\documentclass[%
 aip,
 amsmath,amssymb,
 reprint,%
]{revtex4-1}

\usepackage{graphicx}%
\usepackage{dcolumn}%
\usepackage{bm}%
\usepackage{hyperref}%
\usepackage[utf8]{inputenc} 
\usepackage[english]{babel}
\usepackage{graphicx}
\usepackage{color,epsfig}
\usepackage{amsmath,amssymb}
\usepackage{empheq}

\newcommand{\ffig}[1]{Fig.~\ref{#1}}

\begin{document}

\title{Hopping and emergent dynamics of optical localized states in a trapping potential}

\author{B. Garbin}
\affiliation{Université Paris-Saclay, CNRS, Centre de Nanosciences et de Nanotechnologies, 91120, Palaiseau, France}
\author{J. Javaloyes}
\affiliation{ Departament de Física \& IAC-3, Universitat de les Illes Balears, C/ Valldemossa km 7.5, 07122 Mallorca, Spain}
\author{G. Tissoni}
\author{S. Barland}
\affiliation{Université Côte d'Azur, CNRS, Institut de Physique de Nice, F-06560 Valbonne, France}

\date{Compiled \today}

\begin{abstract}

	The position and motion of localized states of light in propagative geometries can be controlled via an adequate parameter modulation. Here, we show theoretically and experimentally that this process can be accurately described as the phase locking of oscillators to an external forcing and that non-reciprocal interactions between light bits can drastically modify this picture. Interactions lead to the convective motion of defects and to unlocking as a collective emerging phenomenon.

\end{abstract}

\maketitle

\textbf{Localized states can in principle exist at any position in spatially extended nonlinear systems, provided space is invariant by translation. As a consequence, they are expected to move whenever there is a deviation of perfect translational invariance or to diffuse under the influence of noise. A suitable spatial modulation of a parameter may enable pinning of localized states. We propose an experimental and theoretical analysis of pinning optical localized states in a propagative geometry. We show that the pinning of a single localized state can be understood as phase locking of an oscillator. When several localized states coexist along the direction of propagation of light, their mutual interactions are asymmetric: they lack Newton's third law. This leads to a modification of their natural propagation velocity, which in turns influences their un-pinning from the external modulation. The unlocking transition is then a collective phenomenon ruled by the number of interacting elements.}

\section{Introduction}

Several kinds of localized states (LSs) of light have been observed in many propagative experimental configurations including forced nonlinear resonators \cite{temporalcs,herr2014temporal,brasch2016photonic,webb2016experimental,anderson_coexistence_2017,nielsen_coexistence_2019}, subcritical lasers \cite{marconi2014lasing}, or lasers with external forcing \cite{garbin2015topological,PhysRevLett.115.043902}. In spite of their differences they share many properties, in particular the existence of a neutral mode associated to their translation. Thanks to this feature, any perturbation non-orthogonal to this neutral mode will cause motion of LSs \cite{maggipinto2000} and adequately prepared landscapes can be used to trap localized solutions. This has been realized experimentally in many systems, both in the transverse dimension \cite{gutlich2005forcing,bortolozzo2006,positioning06,gutlich2007dynamic} and along propagation \cite{jang2015temporal,jang2016all,javaloyes2016dynamics,camelin_electrical_2016,obrzud_temporal_2017,wang_addressing_2018}. In transverse systems, it has already been noted that the dynamics of a cavity soliton nucleating on and escaping from a trapping position in presence of a drift term can be described as a saddle-node or saddle-loop bifurcations, depending on the depth of the trapping site \cite{PhysRevLett.110.064103,caboche:163901,PhysRevA.80.053814,PhysRevLett.111.233901}. Along the propagation direction, theoretical and numerical studies focussed on the impact of additional forces (inhomogeneities or Raman scattering) on the trapping phenomenon \cite{parra2014effects,hendry_impact_2019}.
In this propagative geometry context, however, most experiments do not offer access to a complete vision of the phenomenon, first due to the difficulty to resolve completely both the fast intra round trip dynamics and the slow evolution over many round trips and second because interactions are in general not considered in the trapping process. Here we show, on the basis of experimental data and analytical calculations, that a unifying view of the trapping process of a single structure is that of phase locking of an oscillator, but also that interactions between LSs can drastically alter their stability in a periodic landscape.

In propagative geometries, due to the fundamentally periodic nature of optical resonators, each dissipative soliton forming along the direction of propagation (e.g. in Kerr fiber ring cavities or semiconductor lasers with feedback) can be considered as an oscillator whose natural frequency is set by the soliton's round-trip time in the resonator. Attempting to tweeze or trap these light bits with a parameter modulation simply consists in trying to lock the phase and frequency of each oscillator to that of an external clock. The detuning between the forcing frequency and the nearest multiple of the inverse of the round-trip time breaks the parity symmetry in a very similar way to a drift force in transverse systems.

In the following we study the unpinning transition of a single dissipative soliton and identify experimentally the signature of the saddle-node on a circle (SNIC) bifurcation. We also analyse in terms of oscillator phases the spatial distribution of LSs in the locked and unlocked regimes, relating their collective motion to the lack of a common clock (diffusive process), but also to their asymmetrical repulsive interaction (see especially \cite{munsberg2020topological} in this same issue). When several LSs are pinned they form a lattice and due to their interactions we observe the propagation of a dislocation in this lattice, \textit{i.e.} a \textit{meta}soliton closely related to supersolitons and soliton Newton craddle predicted theoretically in conservative \cite{novoa2008supersolitons,driben2013newton,ma2016solitons} and dissipative \cite{besse2013pattern,parra2014effects} settings. These observations can be reproduced numerically and the equation describing the unpinning bifurcation of a single soliton can be established analytically. This equation happens to also be very close to the one that describes the formation of solitons in this system, based on a forced oscillator \cite{garbin2015topological}. When many structures are present, we show that beyond simply altering the diffusive process, soliton interactions lead to an anticipated unlocking transition.

\section{Experiment}

\begin{figure}[t]
	\setlength{\unitlength}{\columnwidth}
	\begin{picture}(1,0.7)
		\put(0,0){\includegraphics[width=\columnwidth]{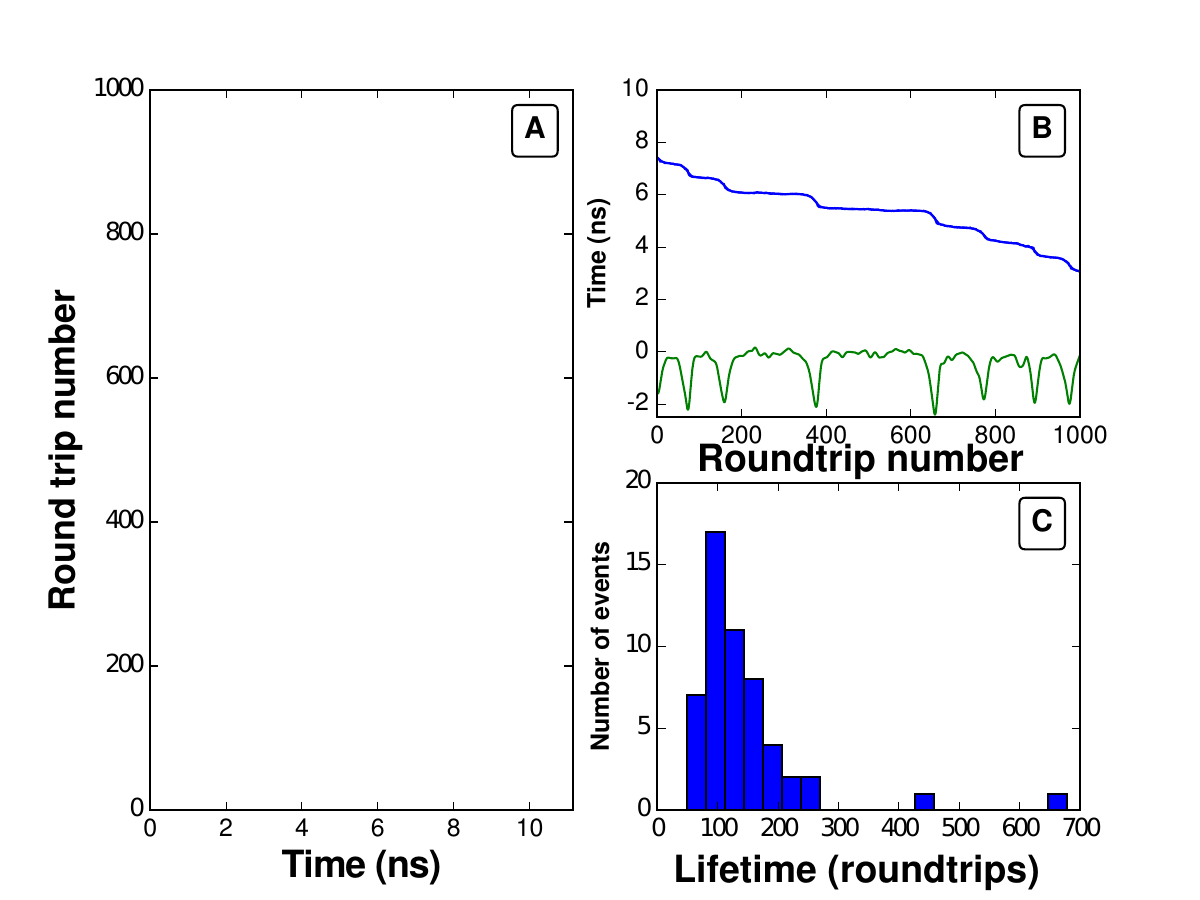}}
		\put(0.15,0.26){\includegraphics[width=3cm]{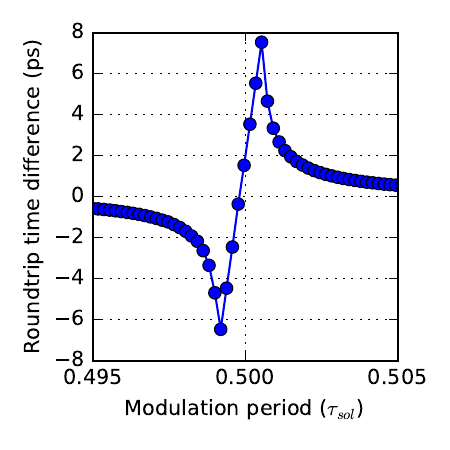}}
		\put(0.32,0.35){\color{red}\circle{0.05}}
	\end{picture}
	\caption{Trajectory of a dissipative soliton in a periodic trap close to the unpinning transition. A) Experimental spatiotemporal diagram in the reference frame of the free running soliton. Inset: numerically obtained locking diagram. In the locked (central) region, the soliton's round trip time deviates from its natural value and tracks the modulation period. B) Position (blue) and velocity (green) of the LS in the reference frame of the trap. C) Histogram of residence time.}
 \label{fig:singleguy}
\end{figure}

The experimental setup is based on a single transverse and longitudinal mode Vertical Cavity Surface Emitting Laser (VCSEL) with optical injection and delayed feedback \cite{garbin2015topological}. The VCSEL is biased at very high current (typically seven times the lasing threshold), ensuring high damping of relaxation oscillations. The injection beam frequency is detuned of about 5~GHz from that of the standalone laser and the amount of injected power (a few percent of the emitted power) is set such that the VCSEL field is phase locked to the external forcing, very close to the unlocking transition. In this regime the laser behaves as an neuron-like excitable system, well described by the Adler equation \cite{coullet1998optical}. A very low reflectivity external mirror (typically 1~\%) is placed in front of the VCSEL so as to define with the latter a compound cavity (typically 2~m long) in which dissipative solitons based on the SNIC phase space structure are stable and can be independently addressed \cite{garbin2015topological} via a Lithium Niobate phase modulator operating on the driving beam \cite{turconi2013control,garbin2013control}. We trap the solitons by driving the Lithium Niobate phase modulator, with an additional sinusoidal signal whose frequency is close enough to a multiple of the inverse of the round-trip time of the soliton in the extended cavity $\tau_{sol}$. In the comoving reference frame, this results in a spatially periodic modulation of the background solution. However (unlike the usual situation in transverse systems) this trap continuously shifts towards one or the other side with respect to the stationary soliton, depending on the detuning between the modulation frequency and the nearest multiple of the inverse round-trip time. The dynamics of a single structure can be analyzed in detail via numerical integration of a set ordinary differential equations with delay \cite{garbin2017interactions}. Depending on the detuning between the phase modulation frequency and the soliton round-trip frequency, the latter can lock to the modulation or drift away from it (see \ffig{fig:singleguy}). 

In \ffig{fig:singleguy}A), the oblique stripes reveal the spatial modulation caused by the trapping modulation. The modulation frequency ($f_{mod}=2.27$~GHz) is slightly higher than 20 times the inverse of the soliton round-trip time and the stripes' slope shows the nonstationarity of the trap due to this detuning. On the inset, we show the numerically observed soliton roundtrip frequency depending on the phase modulation frequency (see also Sec.~\ref{sec:theory} for details). It shows that the soliton round trip time tracks a submultiple of the modulation period (central portion) until it abruptly unlocks, in a way that is compatible with a saddle-node on a circle bifurcation (see \textit{eg} \cite{pedaci2011excitable} for an optomechanical example). The soliton is locked in the central region and unlocked elsewhere. The red circle indicates the experimentally studied parameter region. At the unlocking transition, the soliton is most of the time trapped into meta-stable states from which it randomly jumps away into a neighbouring trap. The direction of the jumps is set by the detuning between the modulation frequency and the closest multiple of the soliton inverse round-trip time. On \ffig{fig:singleguy} B), we plot the position and speed of the soliton in the reference frame of the trap. The position shows the trajectory of an overdamped particle in a tilted periodic potential \cite{coullet:1122}. Correspondingly, the velocity is essentially zero during long periods of time, separated by random spikes all identical to each other which correspond to the random jumps of the particle from one well to the next. The transition between trapping and escaping of the soliton therefore takes place as large amplitude events identical to each other and irregularly distributed in time. This transition (almost infinite period with finite amplitude) is compatible with a SNIC bifurcation in presence of noise which triggers the jumps. We show on \ffig{fig:singleguy}C) the histogram of the particle residence time in a given well. Although the statistical sample is not extremely large (about 50 events in total), this histogram can readily be interpreted in terms of an exponential decay at large times (manifestation of Kramer's law \cite{RevModPhys.62.251}) with a cut-off at short time basically set by the duration of the velocity spikes themselves. This dynamics is perfectly analogue to that of an overdamped particle in a tilted potential in a noisy environment described by the Adler equation\cite{adler1946study,costantini1999threshold}, which also describes the evolution of a forced phase oscillator at the unlocking transition \cite{strogatz2014nonlinear,coullet:1122}.

\begin{figure}[t]
	\setlength{\unitlength}{\columnwidth}
	\begin{picture}(1,0.5)
		\put(0,0){\includegraphics[width=\columnwidth]{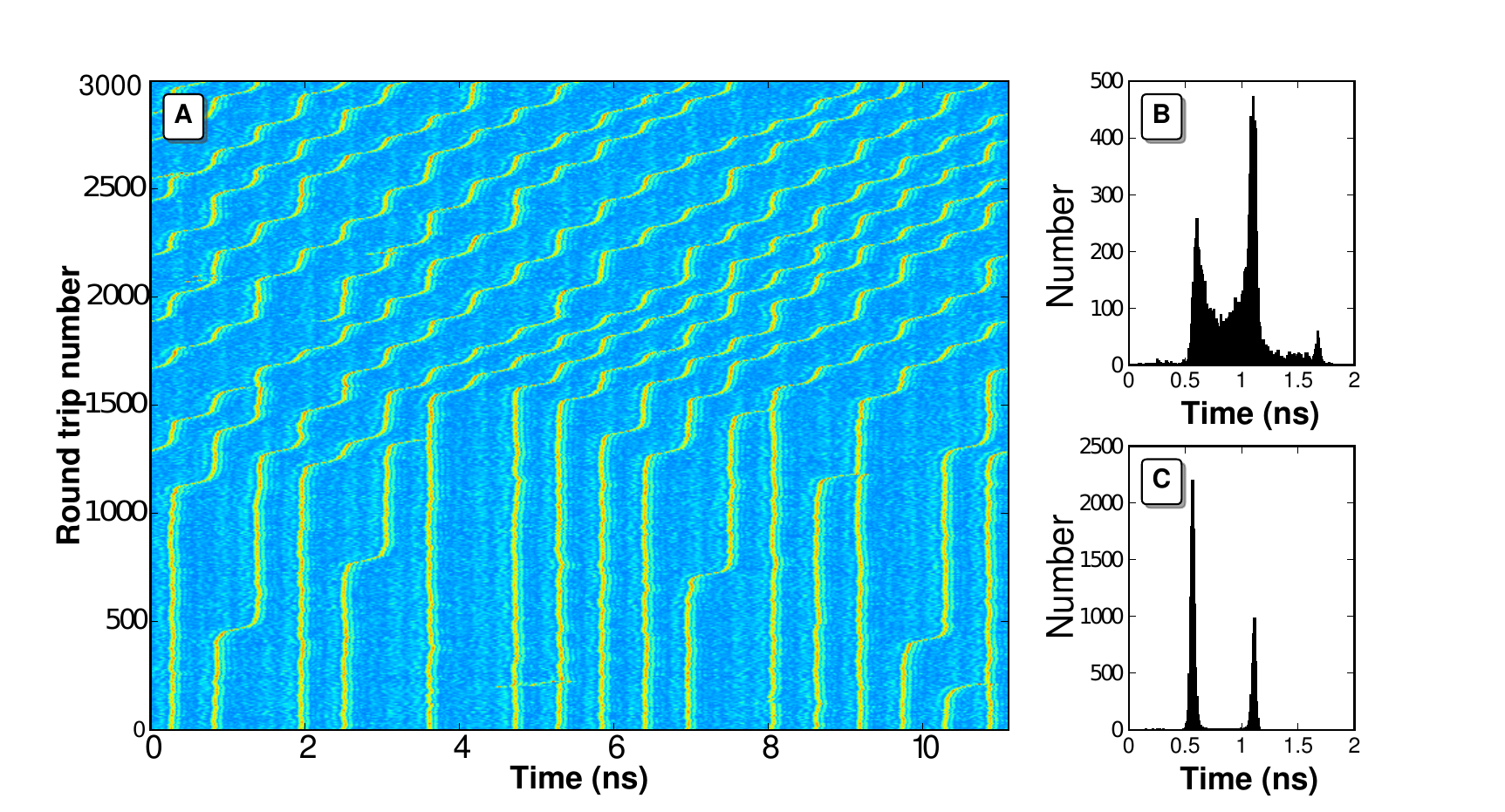}}
		\setlength{\unitlength}{1mm}
		\put(56,7.4){\color{red}\circle{3}}
		\put(49,19.2){\color{red}\circle{3}}
	\end{picture}
  \caption{Unpinning transition with many interacting structures. A) Spatiotemporal diagram
  in the reference frame of the trap. B),C) Histogram of the solitons inter-distances respectively plotted for the last (resp. first) 1000 round-trips.}
 \label{fig:transition}
\end{figure}

The unpinning transition of many solitons is shown on \ffig{fig:transition}. Here we place the observer in the reference frame of the trapping landscape. In the first 1200~round-trips, 16 solitons occupy 16 stable traps out of the 20 available. From time to time some solitons escape randomly and fall into the empty neighbouring trap. In the case of the events occurring at (10.5,250) or (9,1200) the neighbouring trap is already occupied and one of the two solitons vanishes because two solitons can not occupy one single narrow trap whose temporal duration is set by the modulation frequency to $ 1/(2f_{mod})=0.22~$ns. This is due to a repulsive interaction between them \cite{garbin2015topological}, caused by the refractory time following excitable spikes in this injected laser (of the order of  0.4~ns \cite{garbin2017refractory}).

At around round-trip 1500, all solitons progressively start escaping their traps towards the right, each of them following a periodically oscillating trajectory. Considering the solitons as non interacting oscillators, this change of behavior can be seen as an unlocking transition from the common forcing, but this description is very incomplete as we shall see. We show on \ffig{fig:transition}B),C) the histograms of the temporal separations between solitons. On panel C) (trapped regime) two peaks are clearly visible and well separated. Their absolutely minimal width (less than 100~ps) results from all the oscillators being locked: their relative phase is defined to the precision of the trapping potential acting as a master clock. On the second histogram (round-trips 1500 to 2500) the remnants of the peaks are considerably broadened.  Since the oscillators are not locked, their phase with respect to the forcing deterministically drifts (due to the detuning) and  randomly diffuses due to noise. This diffusion is related to the dispersion in the response time of excitable systems near the excitation threshold \cite{garbin2017refractory,garbin2018excitable}. Diffusion of overdamped particles in tilted potential is dramatically enhanced close to the drifting threshold \cite{reimann2001giant}, which explains the very large broadening of the second histogram as well as the emergence of a third peak. However, purely diffusive motion can be expected only when particles do not interact, which is not the case here: the gaps between the first and second (resp. second and third) peak of the histogram tend to close very fast, but the gap before the first peak remains extremely pronounced in the unlocked regime. This gap demonstrates that interactions strongly influence the phase diffusion of the oscillators i.e. the collective motion of dissipative solitons in the trapping landscape. 

\begin{figure}[t]
    \includegraphics[width=1.\columnwidth]{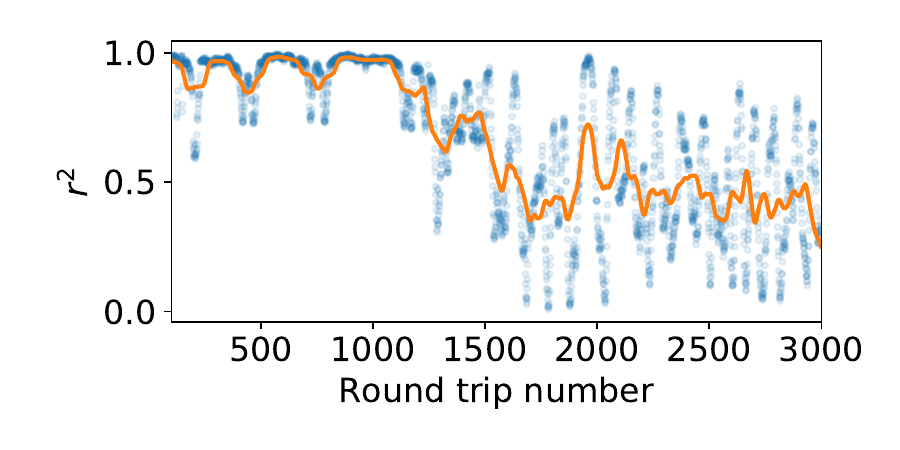}
    \includegraphics[width=1.\columnwidth]{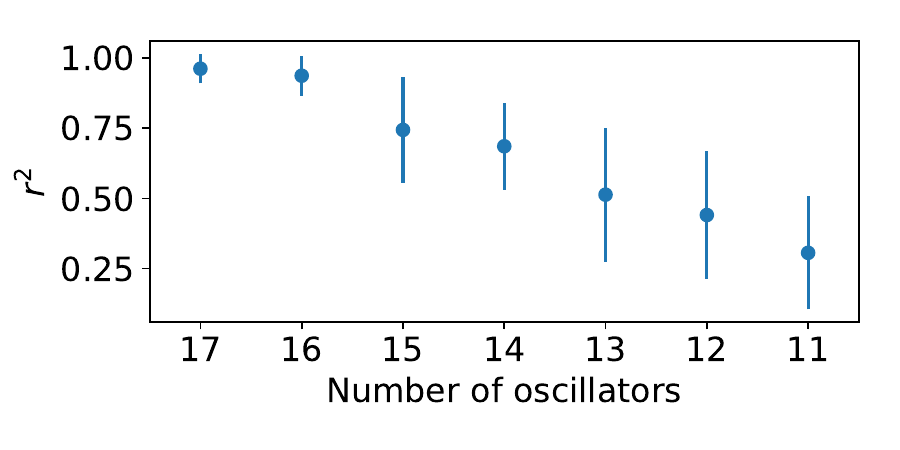}

	\caption{Synchronization to the external forcing. Top: $r^2$ as a function of time. Due to the rather small number of elements to average over, the data is noisy (blue dots) and the curve is better read with smoothing over time (orange line). Bottom: $r^2$ as a function of the number of oscillators. The large error bars are an effect of the small number of oscillators.}
 \label{fig:sync}
\end{figure}

Further insight on the role of interactions can be gained from an analysis of a synchronization indicator as a function of the number of particles. To quantify the overall synchronization of the full population, we measure the square modulus of the Kuramoto order parameter $r^2 = \frac{1}{N^2}[(\sum_j \sin\phi_j)^2+(\sum_j \cos\phi_j)^2]$ where the $\phi_j$ are the phases of each oscillator with respect to the common forcing. This measurement is shown on \ffig{fig:sync}. On the top panel, $r^2$ as a function of time the round trip number shows that a transition taks place about roundtrip 1200, the system switching away from complete synchronization to a much more disordered phase. Interestingly, this transition can also be observed when measuring $r^2$ as a function of the number of oscillators which are present simultaneously. The bottom panel of \ffig{fig:sync} suggests that the system departs from almost perfect synchronization towards a more disordered phase when the number of oscillator decreases below 16. This suggests that the transition is strongly impacted by the interactions between the dissipative solitons.

\begin{figure}[t]
    \includegraphics[width=0.5\columnwidth]{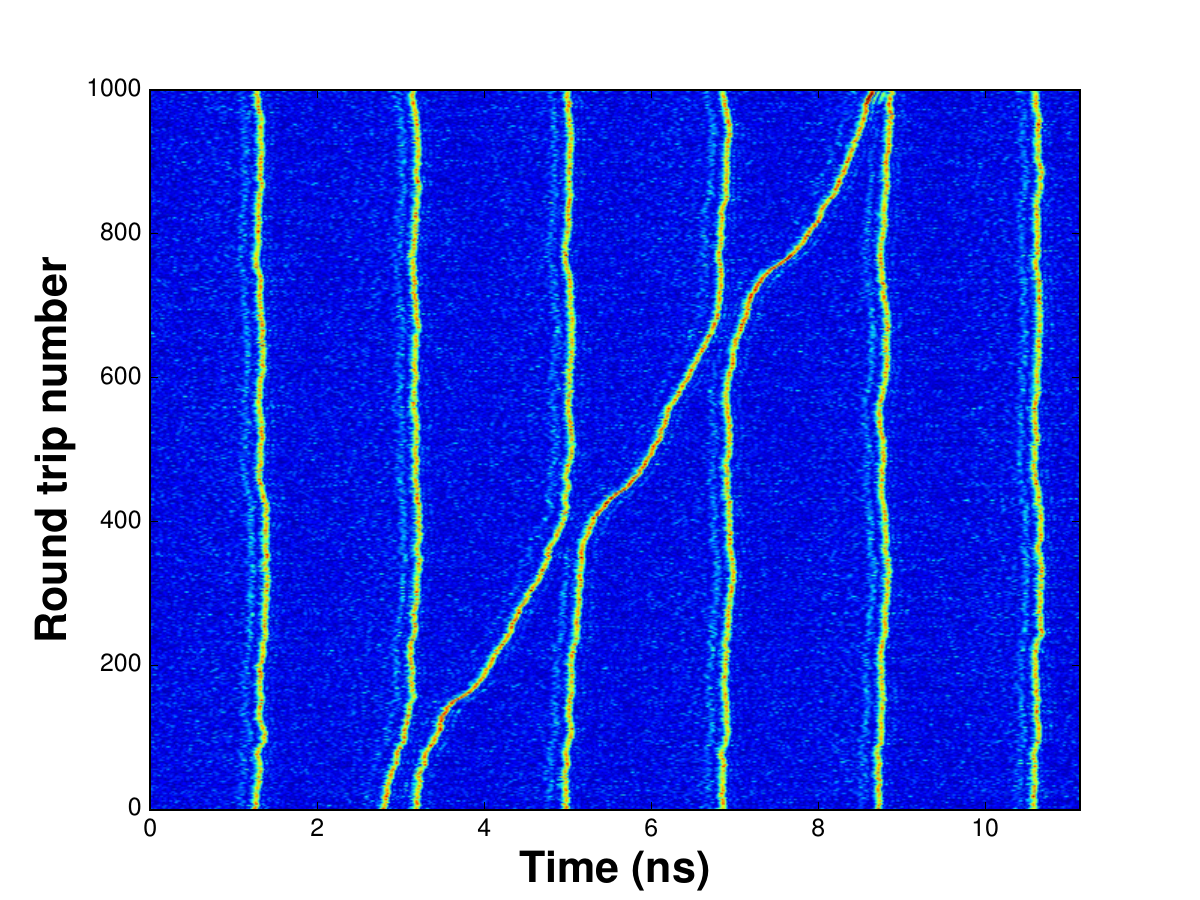}\includegraphics[width=0.5\columnwidth]{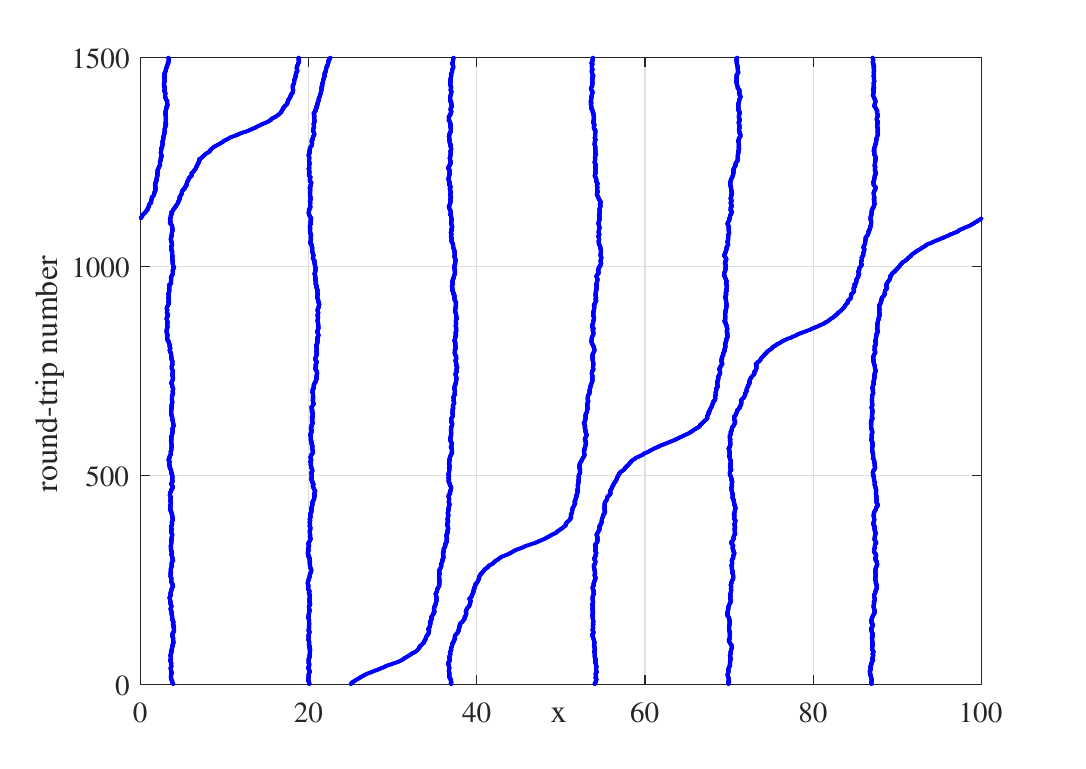}

    \caption{Newton craddle with light bits. Left: experimental observation of seven interacting LS leading to the propagation of a dislocation in the soliton lattice. Right: Numerical simulations of Eq.~\ref{eq:particle} with parameters $\bar{\Delta}=-0.76$, $\psi=0$, $M=0.6$, $\omega=12\pi/L$ and $L=100$, $\sigma=0.1$. }
 \label{fig:meta}
\end{figure}

An extreme example of the impact of interactions is shown on \ffig{fig:meta}. Here we show the trajectories of seven solitons coexisting in a (stationary) six-traps landscape. Here the trap is shallow and repulsive interactions prevent the coexistence of two LS inside one single trap. Upon the arrival of one soliton the other is expelled and falls into the neighboring trap, where the process repeats itself. Thus, the flow of solitons is fundamentally due to their interactions. The disturbance propagating from left to right in \ffig{fig:meta} is analogous to a dislocation in a soliton lattice due to the period mismatch between the lattice and the underlying potential: a \textit{meta}soliton. This is very similar to the supersolitons described theoretically in \cite{novoa2008supersolitons} but here the collisions result from the non-commensurability of the interacting soliton solution with the trap's period.

\section{Theoretical analysis}
\label{sec:theory}

The phase locking of a single structure can be reproduced and analyzed numerically by running simulations of a set of differential equations with the addition of a delayed retroaction term, established in \cite{garbin2017interactions}:

\begin{eqnarray}
	\label{eq:phys}
	\dot{E} & = & \kappa \big[Y(t)+P - (1+i\Theta)E + Ke^{i\phi}E(t-\tau)\big]  \nonumber \\
\dot{P} & = & \Gamma(D)(1+i\Delta(D)) \big[ (1-i\alpha)DE -P\big]  \\
\dot{D} & = & \gamma \big[I_{sl}-D-\frac{1}{2}(EP^*+E^*P)\big] \nonumber 
\end{eqnarray}

where $E$ and $P$ are the slowly varying envelopes of the electric field and of the effective macroscopic polarization,
respectively, $D$ is the excess of carrier density with respect
to transparency. $Y(t)$ is the amplitude of the injected field, that is now time dependent, consisting in a CW field $Y_0$ (taken
real without loss of generality) plus a sinusoidal phase modulation of amplitude $A_m$ and frequency $\omega_m$: $Y(t) = Y_0 \exp(i  A_m sin(\omega_m t))$

$\Theta$ is the frequency mismatch of the injected field with respect to the closest empty cavity resonance (normalised to the cavity decay rate $\kappa$), $\alpha$ is the linewidth enhancement factor, $\tau$ is the external cavity round-trip time, $K$ and $\phi$
are the feedback amplitude and phase, respectively.
The presence of the equation for the macroscopic polarization P allows us to include in the model the asymmetric gain curve typical of semiconductor media, and the functions
$\Gamma (D) = 0.276 + 1.016 D$, $\delta(D) = –0.169 + 0.216 D$ account
for the gain lineshape and peak frequency dependence on the
carrier density $D$. They are obtained by fitting the microscopic susceptibility of the active medium \cite{hachair_cavity_2006,PhysRevA.75.053811,PratiEPJD2010}.

Time is normalized to the dephasing time of the microscopic dipoles $\tau_d$
(typically 100 fs), whereas $\kappa$ and $\gamma$ are defined as $\kappa = \tau_d / \tau_p$ and $ \gamma = \tau_d / \tau_{nr}$, where $\tau_p$ and $\tau_{nr}$ are the cavity photon lifetime
and the non–radiative carrier recombination time. 
Finally, $I_{sl}$ is a parameter related to the pump current normalized in
such a way that the threshold for the solitary laser is $I_{thr} = 1$.

\begin{figure}[t]
    \includegraphics[width=1.\columnwidth]{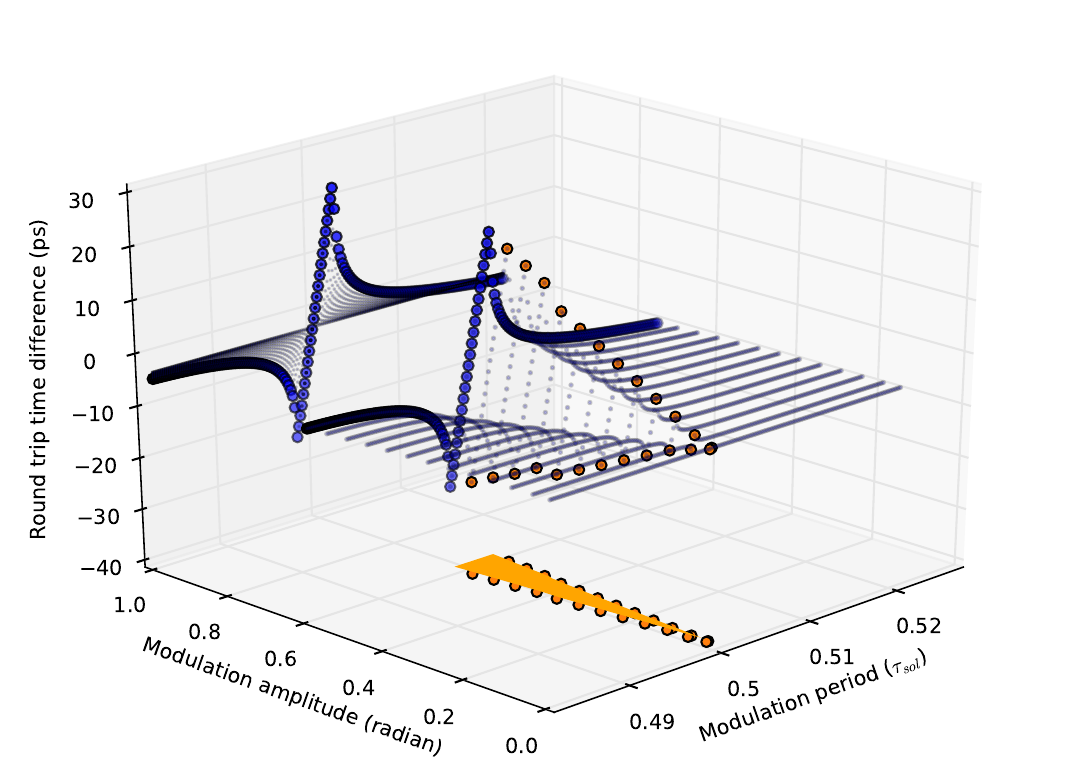}

	\caption{Locking of a soliton to a periodic modulation observed on numerical simulations of the model \ref{eq:phys}. The blue dots correspond to the highest modulation amplitude (0.8~radians) whereas the orange dots show the maximal deviation from the natural period. These points are also projected on the bottom plane of the figure, where they display an Arnold's tongue. Parameters are: $I_{sl}=6$, $\tau=5$~ns, $Y_0=0.01116$, $K=0.00446$, $\Theta=-2.97902$, $\gamma=10^{-4}$, $\kappa=4.10^{-2}$, $\alpha=3$, $\phi=0$.}
 \label{fig:3D}
\end{figure}

On figure \ref{fig:3D}, we show the numerically measured deviation of the single soliton roundtrip time from its roundtrip time in absence of modulation, as a function of the amplitude and the frequency of the trapping potential. 

Here one can clearly recognise that the period of the soliton follows that of the phase modulation over a whole range of frequencies, as shown on the $(y,z)$ projection of the data (also on the inset of \ffig{fig:singleguy}). This range of frequencies over which this locking takes place obviously changes with the amplitude of the forcing, as appears on the horizontal $(x,y)$ plane on \ffig{fig:3D}. The triangular shape of this projection reflects the Arnold's tongue due to locking a periodically propagating dissipative soliton to an external modulation. 

In spite of its insight about the physical origin of the interactions between the dissipative solitons \cite{garbin2017interactions}, the model \ref{eq:phys} is difficult to handle analytically. Thus, we base our following analysis on the simplified model derived in \cite{garbin2015topological} for the phase of the emitted field with respect to that of the driving beam. Assuming that $\alpha$ is the Henry linewidth enhancement factor of the laser and the delayed feedback has a phase $\phi$, we define $\psi=\phi+\arctan\alpha$ to find that the equation governing the evolution of $\theta=\phi_{laser}-\phi_{inj}+\arctan \alpha$ reads 

\begin{eqnarray}
\frac{\partial\theta}{\partial\xi} & = & \bar{\Delta}-\sin\left[\theta+\varphi\left(x\right)\right]+\frac{\partial^{2}\theta}{\partial x^{2}}+\tan\psi\left(\frac{\partial\theta}{\partial x}\right)^{2},\label{eq:SG}
\end{eqnarray}
where we introduced the effective detuning $\bar{\Delta}$ and the 
pseudo-space variable ($x$) representing the local position of the 
light bits within a period. The slow temporal variable ($\xi$) 
contains the residual evolution induced e.g. by interactions between LSs, 
noise and the action of the modulated injected field, see \cite{garbin2015topological} for details.
We assume the latter to be of the following form $\varphi\left(x\right)=m\sin\left(\omega x\right)$.

Considering $\left(\bar{\Delta},\psi,m\right)\ll1$, one can reconstruct
analytically the effective equations of motion of several distant LSs
using a variational approach. We evaluate the interactions
between distant kinks by projecting their residual interactions on the
dual of their neutral translation mode.
When $(\bar{\Delta},\psi,m)=0$, analytical $2\pi$ homoclinic orbits
corresponding to kink and anti-kink solutions of Eq.~\ref{eq:SG} are
known and read $\Theta\left(x\right)=4\arctan\exp\left(x\right).$
When $\left(\Delta,\psi,m\right)\neq\left(0,0,0\right)$ the perturbed
solution takes the form 
\begin{eqnarray}
\theta\left(x,\xi\right) & = & \sum_{j=1}^{N}\Theta\left[x-a_{j}\left(\xi\right)\right]+\cdots\label{eq:Ansatz}
\end{eqnarray}
where the dots represent higher order corrections. Multiplying by the 
adjoint eigenvector of the goldstone (translation) mode and integrating 
over $\mathbb{R}$ allows us to find the following dynamical equations 
for the positions $a_{j}\left(\xi\right)$ 
\begin{eqnarray}
\frac{da_{j}}{d\xi} & = & -\frac{\pi}{4}\left(\bar{\Delta}+2\psi\right)-M\left(\omega\right)\sin\left(\omega a_{j}\right)\label{eq:particle}\\
	& - & F\left(a_{j+1}-a_{j}\right)-F\left(a_{j-1}-a_{j}\right)+\sigma\eta(t)\negthickspace.\nonumber 
\end{eqnarray}
with $M(\omega)=m(\pi \omega^{2}/4)\operatorname{sech}(\pi\omega/2)]$ and $\eta(t)$ is a white Gaussian noise accounting for experimental fluctuations.
The first term in Eq.~\ref{eq:particle} represents the additional
drift imposed by the combined presence of the detuning and of the
gradient squared terms. As such the exact period of the perturbed
orbits is slightly different from the one found in the unperturbed case.
The second term stems from the action of the modulated potential. As
the modulation is averaged over the extent of the LSs, its effective
strength $M\left(\omega\right)$ is a function of the frequency $\omega$,
as expected if one remembers that it actually corresponds to the parametric
forcing of an oscillator. The third and fourth terms represent the
force between nearest neighbors. In deriving Eq.~\ref{eq:particle},
we assumed that the kinks are not too close, and that the interaction
terms are small, so that we can safely neglect second order additional
coupling arising via the inertia terms. The effective force $F$ contains
interactions mediated by all the terms of Eq.~\ref{eq:SG}, and, 
although cumbersome, the expression
of $F$ can be found analytically. However, as soon as the distance
between the kinks is larger than a few times their typical width,
one can use an asymptotic approximation that simply reads 
\begin{eqnarray}
F_\psi\left(\delta\right) & = & \left(4\operatorname{sgn}\left(\delta\right)+6\pi\psi\right)\exp\left(-\left|\delta\right|\right),
\end{eqnarray}
with $\delta$ the distance between consecutive LSs. 

Distant LSs interact via their exponentially decaying tails.
While many salient examples of Solitons are even functions, as e.g 
solutions of the Nonlinear Schr\"odinger, Ginzburg-Landau or Lugiato-Lefever 
equations, the left and right exponential tails are not necessarily identical (see \textit{eg}\cite{Wang:17}). 
As such, the interactions between LSs may not be reciprocal and their dynamics may disobey the action-reaction principle, as demonstrated recently in passively 
mode-locked lasers \cite{javaloyes2016dynamics}. This is also the case here.
This fact is not inconsistent with the parity symmetry of Eq.~\ref{eq:SG}.
As we have two solutions branches, the symmetry only maps the kink over 
the anti-kink solution while both solutions are neither even nor odd in general.

From Eq.~\ref{eq:particle} in the case of a single LS, we can clearly
see the possibility of an Adler Locking-Unlocking transition (SNIC bifurcation)
for the LS drifting speed depending on the precise value of $M$ as compared
to the two critical values $M_{c}=\pm \frac{\pi}{4} \left| \bar{\Delta}+2\psi \right|$.
The system possesses a single stable (and an unstable) equilibrium
point, provided that $|M|\ge |M_{c}|$. In this case, the position evolves
into a washboard potential that exhibits for each period a minimum and a maximum corresponding 
to each of these fixed points. 
Sufficiently close to the unlocking transition, excitability is found in the motion of solitons in the trap and 
noise can generate random sequence of excursions where the LS ``falls'' 
from one weakly stable potential minimum toward the next, as depicted 
experimentally in Fig.~\ref{fig:singleguy}. In the presence of an external 
force field, we show numerical simulations in \ffig{fig:meta}b) demonstrating 
the interplay between the action of the modulation potential and the repulsive 
forces between nearest neighboring solitons, which result, as in the experiment, 
in the propagation of a meta-soliton.

\begin{figure}[t]
	\includegraphics[width=1.\columnwidth]{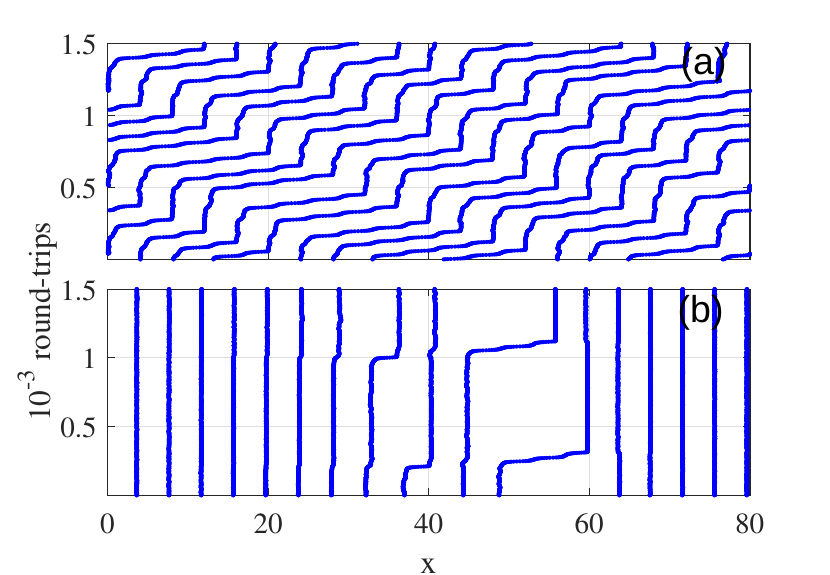}
	\caption{Impact of interactions on the collective motion in the phase model. Left: Numerical observation of 16 LSs stably trapped (bottom) and 11 LSs collectively drifting (top). Both situations coexist with identical parameters in model \ref{eq:particle} and only differ in the number of light particles. Parameters are $\bar{\Delta}=-0.47$, $\psi=0.2$, $M=0.052$, $\omega=40\pi/L$, $L=80$ and $\sigma=0.018$%
    \label{fig:collective}}
    \end{figure}

The fundamentally collective phenomenon depicted in \ffig{fig:transition} can be explained in the framework presented in Eq.~\ref{eq:particle}. Interactions between LSs renormalize their drifting speed which becomes a function of $N$, the number of LSs. As the drifting speed (or equivalently, the repetition period of the oscillator) of $N_1$ LSs is slightly different from that of $N_2$ LSs, one foresees that a periodic solution with $N_1$ LSs could be locked to the frequency of the external modulation while another solution with $N_2$ LSs would not, as is exemplified on \ffig{fig:transition} and \ffig{fig:meta}. While global coupling between LSs due to the presence of a slow variable, such as e.g. thermal effects, is  commonly found in the framework of spatial and temporal dissipative solitons, this effect is absent from the simple modelling presented by Eq.~\ref{eq:particle}. Yet, a variation of the drift speed as a function of N is still fundamentally present, as a consequence of the non-reciprocity of the interactions. Without this non-reciprocity, the various solutions would not drift at all. Starting from a state with $N_1$ LSs, the ensemble may pass from a globally locked state to a drifting one as shown on \ffig{fig:collective}a,b) if the number of LS changes sufficiently. As in the case of the meta-soliton, interactions are therefore fundamentally at the origin of the dynamics.

\section{Conclusion}

In conclusion, on the basis of strong experimental evidence and analytical calculations, we have provided a unifying view of the trapping and escape processes of trapped solitons in the very simple terms of locking of oscillators, shedding new light on related experiments and bridging the gap with the extensive literature of transversally extended systems. We have demonstrated the impact of dissipative solitons interactions on their collective motion in propagative geometries, relating it to a dependence of their round-trip frequency on their number via their non-reciprocal interactions. As a particular example of collective dynamics in a mesoscopic ensemble of light bits, we have also demonstrated experimentally the existence of a defect propagating in a lattice of dissipative optical solitons, \textit{i.e.} of a \textit{meta}soliton. Also, the experimental observation of a globally locked state to a disordered phase when the number of interacting oscillators change suggests that this platform could be interesting in the more general framework of synchronization in complex systems. 

\section{Data availability statement}

The data that support the findings of this study are available from the corresponding author upon reasonable request.

\begin{acknowledgments}
J.J. acknowledge the financial support of the MINECO Project MOVELIGHT (PGC2018-099637-B-100 AEI/FEDER UE). S.B., G.T. and B.G. acknowledge the support from Région Provence Alpes Côte d'Azur through grant number DEB 12-1538.
\end{acknowledgments}

\bibliography{references}

\end{document}